\documentclass[%
 reprint,
 amsmath,amssymb,
 aps,prd,nofootinbib,showpacs,twocolumn,
 mathrsfs,amsfonts,dsfont
]{revtex4-1}
\usepackage{float}
\usepackage{graphicx}
\usepackage{dcolumn}
\usepackage{bm}
\usepackage{wasysym}
\usepackage{verbatim}
\usepackage{color}
\usepackage[linktocpage]{hyperref}
\usepackage{hyperref}
\usepackage{fullpage}
\usepackage{ dsfont }
\usepackage{braket}
\usepackage{scrextend}
\usepackage{wrapfig}

\newcommand{\lsim}{\mathrel{\hbox{\rlap{\lower.55ex\hbox{$\sim$}} \kern-.3em \raise.4ex \hbox{$<$}}}}
\newcommand{\gsim}{\mathrel{\hbox{\rlap{\lower.55ex\hbox{$\sim$}} \kern-.3em \raise.4ex \hbox{$>$}}}}

\newcommand{\mpl}{m_{\mbox{\tiny{Pl}}}}
\newcommand{\Beq}{\begin{equation}\begin{aligned}}
\newcommand{\Eeq}{\end{aligned}\end{equation}}

\makeatletter
\def\l@subsubsection#1#2{}
\makeatother

\begin{document}

\title{Gravitational perturbations from oscillons and transients after inflation}
\author{Kaloian D. Lozanov${}^1$ and Mustafa A. Amin${}^2$}

\affiliation{${}^1$Max Planck Institute for Astrophysics, Karl-Schwarzschild-Str. 1, 85748 Garching, Germany\\${}^2$Physics \& Astronomy Department, Rice University, Houston, Texas 77005-1827, U.S.A.}

\date{\today}
\begin{abstract}
We study the scalar and tensor perturbations generated by the fragmentation of the inflaton condensate into oscillons or transients after inflation, using nonlinear classical lattice simulations. Without including the backreaction of metric perturbations, we find that the magnitude of scalar metric perturbations never exceeds a few $\times 10^{-3}$, whereas the maximal strength of the gravitational wave signal today is $\mathcal{O}(10^{-9})$ for standard post-inflationary expansion histories. We provide parameter scalings for the $\alpha$-attractor models of inflation, which can be easily applied to other models. We also discuss the likelihood of primordial black hole formation, as well as conditions under which the gravitational wave signal can be at observationally interesting frequencies and amplitudes. 

Finally, we provide an upper bound on the frequency of the peak of the gravitational wave signal, which applies to all preheating scenarios.
\end{abstract}

\maketitle

\tableofcontents

\section{Introduction}

\begin{figure}[t] 
   \centering
   \includegraphics[width=2.5in]{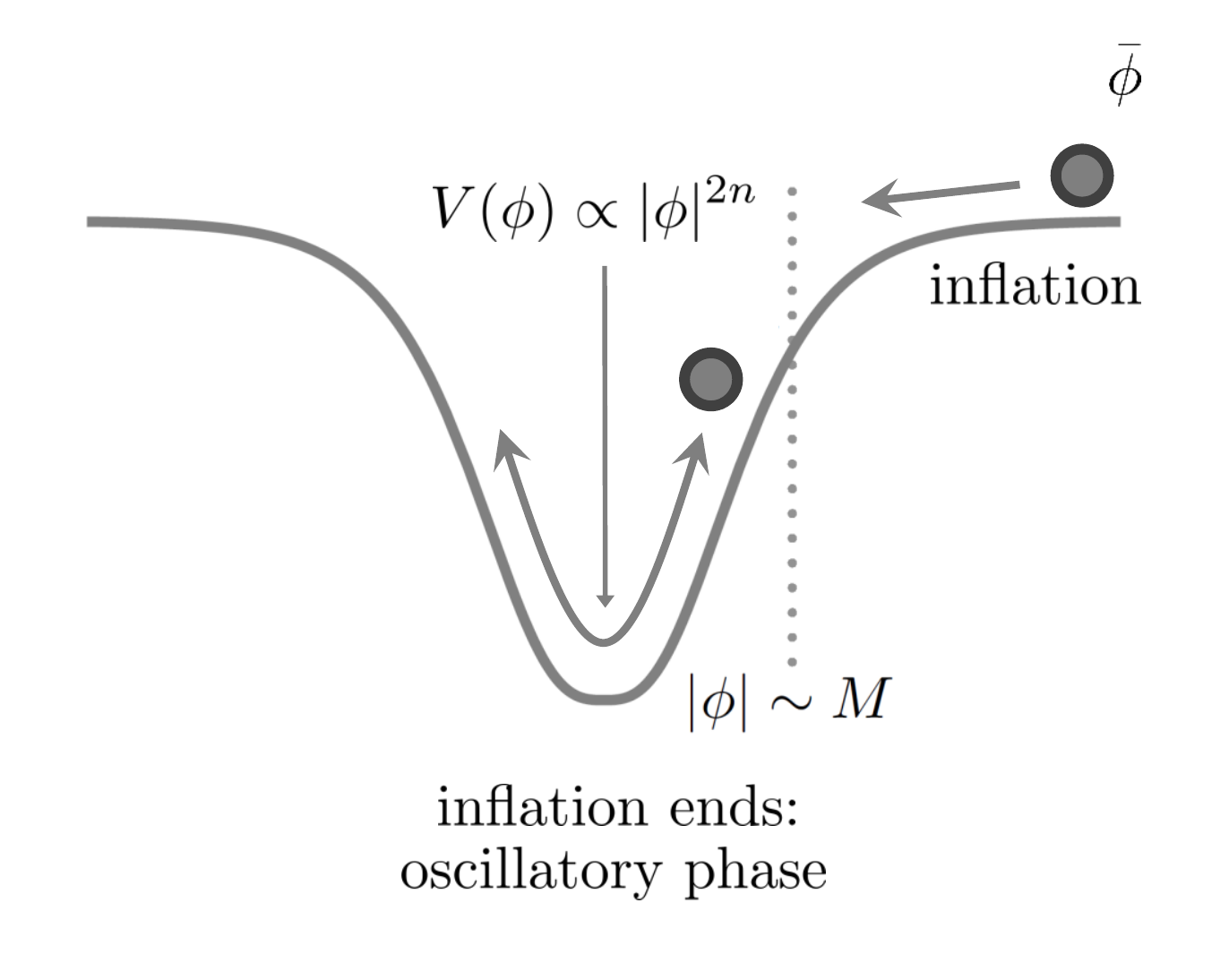}
   \caption{The generic shape of the inflaton potential, $V(\phi)$. During inflation the inflaton rolls slowly along one of the plateaus, $|\phi|>M$, towards the minimum. Inflation ends when the inflaton enters the central power-law region and begins to oscillate about the bottom of $V(\phi)$.}
   \label{fig:Potential1}
\end{figure}

Measurements of the CMB anisotropies \cite{Ade:2015xua,Ade:2015lrj,Array:2015xqh,Akrami:2018vks,Aghanim:2018eyx,Akrami:2018odb} have ruled out to a high statistical significance single-field inflation driven by a monomial inflaton potential \cite{Starobinsky:1980te,PhysRevD.23.347,Sato:1980yn,LINDE1982389,Albrecht:1982wi}. 
A broad class of inflaton potentials which fit the observational constraints, feature a power-law minimum and a plateau-like region away from it, where slow-roll inflation is realised, see Fig. \ref{fig:Potential1}. 

Such potential profiles can have important consequences for reheating -- the period after inflation where the energy of the inflaton is transferred to the Standard Model degrees of freedom, the universe thermalizes and becomes radiation dominated, setting the scene for BBN (for reviews on reheating, see \cite{Boyanovsky:1996sv,Bassett:2005xm,Frolov:2010sz,Allahverdi:2010xz,Amin2014}). The nonlinear effects related to the shape of the inflaton potential are most prominent when the couplings of the inflaton to other fields are suppressed with respect to its own self-interactions, $V(\phi)$, during the initial oscillatory stage of reheating, see Fig. \ref{fig:Potential1}. In this case the parametric resonance coming from the inflaton condensate oscillations leads to the exponential amplification of its own fluctuations (called {\it self-resonance}), as opposed to the standard reheating scenarios where the amplification of daughter fields is assumed \cite{Dolgov:1989us,PhysRevD.42.2491,Kofman:1994rk,Shtanov:1994ce,Kofman1997}. 

The nonlinear dynamics following efficient self-resonance and the backreaction of inflaton fluctuations can be quite rich in potentials which flatten away from the minimum. If the minimum of the potential is quadratic, long-lived pseudo-solitonic objects called oscillons \cite{Bogolyubsky:1976yu,Gleiser:1993pt,Copeland:1995fq,Amin:2010jq,Amin:2013ika,Mukaida:2016hwd} are produced in abundance \cite{Amin:2010xe,Amin:2010dc,Amin:2011hj, Gleiser:2011xj}. In this case, we get a matter dominated equation of state for the universe following inflation. For a power-law (but non-quadratic) minimum in the potential, short-lived pseudo-solitonic objects called transients are produced, followed by a transition to a radiation dominated state of expansion \cite{Lozanov:2016hid,Lozanov:2017hjm}. 

This work can be considered as a follow-up to \cite{Lozanov:2016hid,Lozanov:2017hjm}, where we studied the details of self-resonance and nonlinear inflaton dynamics, and its implications for the post-inflationary expansion history of the universe. Here we focus on the gravitational effects of the nonlinear field dynamics. We concentrate on the cases where oscillons or transients are copiously produced. In this regime, the plateau as well as the central minimum of $V(\phi)$ are relevant for the field dynamics. 

If only the power-law region around the central minimum is probed by the inflaton after inflation, pseudo-solitonic objects do not form copiously. In this case the oscillating condensate can still be destroyed slowly due to its self-interactions \cite{Khlebnikov:1996mc,Micha:2002ey,Micha:2004bv,Lozanov:2016hid,Lozanov:2017hjm} or its own gravity \cite{Easther:2010mr}, sourcing metric perturbations as shown in \cite{Khlebnikov:1997di} and \cite{Easther:2010mr}, respectively. We do not consider this regime here.

In the wake of the exciting aLIGO-Virgo detection of the black-hole binary mergers \cite{Abbott:2016blz} and their implications for primordial black holes (PBHs) abundance as well as their contribution to dark matter \cite{Bird:2016dcv,Clesse:2016vqa}, it is natural to ask if oscillons or transients can seed such PBHs. A recent study invoking statistical arguments \cite{Cotner:2018vug} argued that the inevitable longterm gravitational collapse of an accidentally overdense cluster of oscillons can lead to the formation of PBHs. But can individual oscillons or transients generate sufficiently strong gravitational fields to trigger the collapse of individual objects and the formation of PBHs? We address this question by studying the scalar metric perturbations sourced by oscillons and transients in a cosmological setting at the end of inflation. We employ classical lattice simulations which capture the nonlinear evolution of the inflaton field, but ignore the backreaction of metric perturbations on the field dynamics (which is justified by the small gravitational fields we find, and the few {\it e}-folds of evolution that we consider).

We note that the formation of light PBHs through oscillon collisions \cite{Konoplich:1999qq,Khlopov:1999ys,Rubin:2000dq,Amin:2014fua} and their evaporation through Hawking radiation can provide additional constraints on reheating or another channel for it \cite{Carr:2009jm}. The consequences of the fragmentation of the inflaton condensate for PBHs formation after inflation have been investigated in the context of interacting theories \cite{GarciaBellido:1996qt,Green:2000he,Bassett:2000ha,Hidalgo:2011fj,Torres-Lomas:2014bua,Suyama:2004mz} and in some self-interacting models \cite{Cotner:2018vug,Suyama:2006sr}, but never for oscillons and transients with lattice simulations. 

The formation of oscillons after inflation is also known to source gravitational wave (GW) backgrounds \cite{Zhou:2013tsa,Antusch:2016con,Antusch:2017flz,Antusch:2017vga,Amin:2018xfe,Kitajima:2018zco,Liu:2018rrt}. In this work we compare and contrast the GWs sourced by oscillons and transients and provide some simple generic scalings for the frequency and the energy of the maximally excited tensor modes, which could be used by future experiments (for reviews on stochastic GW backgrounds from reheating and future probes see, e.g., \cite{Easther:2006vd,Easther:2006gt,GarciaBellido:2007af,Dufaux:2007pt,Dufaux:2008dn,Dufaux:2010cf,Adshead:2018doq,Bartolo:2016ami,Figueroa:2017vfa,Caprini:2018mtu,Bartolo:2018qqn}). 

We also provide a {\it generic upper bound} on the frequency of the GW peak, which applies to {\it all preheating models}, including both perturbative and non-perturbative particle production, nonlinear dynamics of the inflaton and/or spectator fields.

The paper is organized as follows. In Section \ref{sec:SelfRes} we establish notation, present the fiducial model and review resonant particle production and the subsequent nonlinear dynamics of a self-interacting inflaton after inflation. Our analytical predictions and lattice calculations for the sourced metric perturbations by the oscillons and the transients are given in Section \ref{sec:GrEfAnEst}. In Section \ref{sec:GWs}, we present the analytic estimates and numerical results for gravitational waves sourced by the nonlinear dynamics. We discuss the likelihood of PBHs formation and the detection of GWs from oscillons and transients in Section \ref{sec:Disc} and give our concluding remarks in Section \ref{sec:Conclusions}.
\section{Self-resonance after inflation}
\label{sec:SelfRes}

The inflationary model we consider is specified by the action
\Beq
S=\int d^4x\sqrt{-g}\left[-\frac{\mpl^2}{2}\mathcal{R}+\frac{\partial_{\mu}\phi\partial^{\mu}\phi}{2}-V(\phi)\right].
\Eeq
The focus of this paper is on the gravitational effects of nonlinear inhomogeneites of the scalar field which are in turn generated by self-interaction terms in $V(\phi)$. Hence, we ignore the couplings to other fields for this paper. 

We work in a perturbed Friedmann-Robertson-Walker spacetime. In the Newtonian gauge, the line element is given by
\Beq
\label{eq:PertMetric}
ds^2
=a^2(\tau)&\left[(1+2\Phi)d\tau^2-(1-2\Phi)\delta_{ij}dx^i dx^j\right]\,\\
&\qquad\qquad\qquad\qquad+a^2(\tau)h_{ij}dx^idx^j\,.
\Eeq
Since we consider a single-field model of inflation, we keep only one scalar metric perturbation, $\Phi$, and neglect the vector ones. We also assume that the anisotropic stress is negligible, which is not true in the nonlinear regime, but we will argue that this assumption will not affect our conclusions. The transverse traceless 3-tensor perturbation, $h_{ij}$, describes the GWs. 
\subsection{Inflaton potentials}

We concentrate on observationally-consistent $V(\phi)$ \cite{Ade:2015xua,Ade:2015lrj,Array:2015xqh,Akrami:2018vks,Aghanim:2018eyx,Akrami:2018odb}. Their salient features are a power-law dependence near the central minimum, $\propto|\phi|^{2n}$, for $|\phi|<M$, and a flattening away from it, see Fig. \ref{fig:Potential1}. To be in agreement with CMB constraints, slow-roll inflation is realized in the flat regions. For concreteness, we consider
\Beq
V(\phi)=\Lambda^4\tanh^{2n}\left(\frac{|\phi|}{M}\right)\,.
\Eeq
This parametrization belongs to the $\alpha$-attractor T-models \cite{Kallosh:2013xya,Kallosh:2013hoa,Kallosh:2013yoa,Kallosh:2013wya,Kallosh:2013maa,Galante:2014ifa,Linde:2015uga,Carrasco:2015rva,Carrasco:2015pla,Roest:2015qya,Scalisi:2015qga,Kallosh:2016gqp}, with $M=\sqrt{6\alpha}\mpl$. The observational bounds on $r$ require $M<10\mpl$. For greater values of $M$ slow-roll inflation continues as the inflaton leaves the plateau and enters the central power-law region, generating unacceptably large $r$. In the following, we consider only sufficiently small $M$, for which inflation ends soon after the inflaton exits the flat region. As far as the other parameters are concerned, $n$ is free and
\Beq
\label{eq:Lambda4}
\Lambda^4=\frac{3\pi^2A_{\rm s}}{N_{\star}^2}M^2\mpl^2\,,
\Eeq
where the measured amplitude of the curvature power spectrum is $A_{\rm s}\approx2.2\times10^{-9}$ \cite{Ade:2015xua,Aghanim:2018eyx} and the co-moving pivot scale leaves the horizon $N_{\star}=60$ {\it e}-folds of expansion before the end of inflation. For the $M$ range we are interested in, the tensor-to-scalar ratio is
\Beq
r=\frac{2}{N_{\star}^2}\frac{M^2}{\mpl^2}\,,
\Eeq
manifesting the attractor nature of this class of models -- upon decreasing $M$, $r$ is attracted to low values. We also define an effective mass squared, $m^2\equiv \lim\limits_{\phi/M\rightarrow0}\phi^{-1}\partial_{\phi}V$, as
\Beq
m^2=2n\Lambda^2\left(\frac{\Lambda}{M}\right)^2\left(\frac{\phi}{M}\right)^{2(n-1)}\,.
\Eeq
It determines the effective frequency of the condensate oscillations immediately after inflation, described in the following section.

\subsection{Resonant particle production}
\label{sec:ResPartProd}

The end of slow-roll inflation is followed by an oscillatory phase, during which the inflaton condensate oscillates about the central power-law minimum of its potential, see Fig. \ref{fig:Potential1}. The periodic evolution and the nonlinearities in $V(\phi)$ can lead to resonant amplification of the small perturbations in $\phi$ \cite{Lozanov:2016hid,Lozanov:2017hjm}. Depending on the value of the scale $M$, there are two distinct instability regimes.

\subsubsection{Broad resonance}

For $M\ll\mpl$, the amplitude of the inflaton oscillations decays very slowly due to the expansion of the universe,\footnote{The decay rate is comparable to $H$.}\footnote{For a recent study of the instability regime, inlcuding the effects from the expansion of the universe, see Ref. \cite{Koutvitsky:2018afp}} over many oscillatory time scales. The condensate periodically probes the transition region near $|\phi|\sim M$. This leads to a strong instability in a broad wavelength range of sub-horizon inflaton perturbations. Within less than an {\it e}-fold of expansion, their exponential growth, known as {\it broad resonance}, makes mode-mode couplings, as well as backreaction on the condensate, non-negligible.

\subsubsection{Narrow resonance}

When $M\sim\mpl$, the initial Hubble-induced decay rate of the amplitude of condensate oscillations is comparable to the oscillation frequency. The inflaton is redshifted very quickly towards the bottom of its potential, $|\phi|\ll M$, where the condensate oscillations become lightly damped. There is no time for broad resonance to be established. However, for $n>1$, the small amplitude oscillations at late times are in an intrisically nonlinear region, $V(\phi)\propto|\phi|^{2n}$. They lead to a slow, but steady growth of a narrow wavelength range of sub-horizon inflaton perturbations. This {\it narrow resonance} develops over several {\it e}-folds of expansion, eventually leading to substantial backreaction effects. If $n=1$, no resonant instabilities are present at all. \footnote{If we take into account the {\it gravitational} self-interactions of an inflaton condensate, oscillating about a quadratic minimum, fragmention does eventually occur \cite{Easther:2010mr}, but its typical time scale is much longer than the oscillation time-scale.}

\subsection{Nonlinear dynamics}

Shortly after the unstable $\phi$ perturbations backreact on the condensate, nonlinear dynamics ensues \cite{Lozanov:2016hid,Lozanov:2017hjm}. It can be divided into three different types, depending on the kind of the preceding resonant instabilities (see Section \ref{sec:ResPartProd}), as well as the value of the power $n$.

\subsubsection{Oscillons}
\label{sec:Osc}

The backreaction after broad resonance about a quadratic minimum, i.e., when $M\ll\mpl$ and $n=1$, leads to the formation of oscillons. They are quasi-stable and quasi-spherical, highly overdense field configurations, with typical density contrast $\delta\equiv\rho/\bar{\rho}\geq\mathcal{O}(10)$. Within the objects, $\phi$ has an approximately spherically symmetric profile, oscillating at an angular frequency close to (but slightly smaller than) the effective mass, $m$. Their typical physical size, $R\sim10m^{-1}$, is much smaller than the horizon scale. They can survive for at least millions of oscillations \cite{Salmi:2012ta,Ibe:2019vyo}, which can be equivalent to $\mathcal{O}(10)$ {\it e}-folds of expansion. Since the $\phi$ oscillons dominate the energy budget and behave as pressureless dust, they can lead to a long period of matter-dominated state of expansion with an effective equation of state $w\approx0$.

\subsubsection{Transients}
\label{sec:Trans}

When $M\ll\mpl$, but $n>1$, the backreaction after the broad resonance leads to the formation of objects which resemble oscillons, but are much less stable. We refer to these highly nonlinear, but ephemeral lumps of energy density as transients \cite{Lozanov:2016hid,Lozanov:2017hjm}. Their properties are very similar to those of oscillons if one uses $m=m(\phi=M)$ as the transients effective mass. The only difference is that they survive for tens of oscillations only, which is $\ll\mathcal{O}(1)$ {\it e}-folds of expansion in contrast to the much longer lifetime of oscillons. After their decay, $\phi$ enters into a relativistic turbulent regime, see Section \ref{eq:TurbInfl}, with $w\approx1/3$.

\subsubsection{Turbulent inflaton}
\label{eq:TurbInfl}

The backreaction following narrow resonance, occuring for $M\sim\mpl$ and $n>1$, does not lead to the formation of non-trivial field configurations like oscillons or transients. But still, the inflaton enters a highly inhomogeneous state, which can persist for many {\it e}-folds of expansion. It involves a fragmented density configuration on very small scales, with a continuous slow fragmentation to even shorter scales. The latter can be interpreted as a slow momentum upscatter due to the weak nonlinear self-interactions. The kinetic and gradient energies are approximately equal to each other and much greater than the self-interaction energy, yielding $w\approx1/3$.\footnote{This behavior appears to persist when couplings to additional light fields are introduced \cite{Maity:2018qhi}.} We note that this inhomogeneous state of $\phi$ is also observed after the decay of the transients for $M\ll\mpl$ and $n>1$ \cite{Lozanov:2016hid,Lozanov:2017hjm}. The one-point PDF of the density contrast, $\delta$, fits well a lognormal distribution, see Section \ref{sec:Transients}, which is a compelling indication for scalar field turbulence \cite{Micha:2002ey,Micha:2004bv,Frolov:2008hy,Frolov:2010sz}. On the other hand, when $n=1$, $\phi$ does not evolve into this state (due to $V(\phi)$) since it either forms long-lived oscillons (for $M\ll\mpl$) or remains approximately homogeneous (for $M\sim\mpl$).

\section{Scalar gravitational perturbations}
\label{sec:GrEfAnEst}

The post-inflationary resonant production of inflaton particles, caused by the nonlinear self-interactions of the inflaton field, can lead to the emergence of oscillons and transients as described in the previous section. In this section: (A) We provide analytic estimates for the Newtonian potential, $\Phi$, induced by the formation of such post-inflationary oscillons and transients. (B) We carry out detailed lattice simulations of the nonlinear field dynamics, calculate $\Phi$ from the simulations, and compare our analytic predictions with the results from these simulations. 

\subsection{Analytic estimates}

\subsubsection{Linear regime}

When the inflaton field starts oscillating coherently at the end of inflation, one can work to first order in field perturbations. The inflaton can be split into a time-dependent background component and space-dependent perturbations, $\phi(x,\tau)=\bar{\phi}(\tau)+\delta\phi(x,\tau)$. The generalized Poisson equation, following from a combination of the linearized Einstein equations \cite{Bassett:2005xm}, determines the evolution of the (non-dynamical) scalar metric perturbation, $\Phi$, see eq. \eqref{eq:PertMetric}:
\Beq
\frac{\nabla^2 \Phi}{a^2}=\frac{\delta\rho_{\rm m}}{2\mpl^2}\,,
\Eeq
where $\delta\rho_{\rm m}=\delta\rho-3\mathcal{H}\bar{\phi}'\delta\phi/a^2$ is the co-moving density perturbation, $\delta\rho=\delta T^0_{\,\,\,0}=\bar{\phi}'\delta\phi'/a^2+\partial_{\bar{\phi}}\bar{V}\delta\phi$ is the inflaton density perturbation and $'\equiv\partial_{\tau}$. Since the condensate oscillations are lightly damped, on average $3\mathcal{H}|\bar{\phi}'|\ll a^2\partial_{\bar{\phi}}\bar{V}$. Similarly, on sub-horizon scales $3\mathcal{H}|\delta\phi|\ll|\delta\phi'|$. Hence, to a good approximation, for $k\gg\mathcal{H}$,
\Beq
\label{eq:GenPoisson}
\tilde{\Phi}_{\bf{k}}\approx-\frac{3}{2}\left(\frac{\mathcal{H}}{k}\right)^2\frac{\delta\tilde{\rho}_{\bf{k}}}{\bar{\rho}}\,.
\Eeq
Since the resonant instabilities occur on sub-horizon scales, i.e., $\delta\tilde{\rho}_{\bf{k}}/\bar{\rho}$ grows for $k\gg\mathcal{H}$, the scalar metric perturbation can grow, but remains small, $\Phi\ll1$, due to the $\mathcal{H}/k$ suppression factor in eq. \eqref{eq:GenPoisson}.\footnote{On super-horizon scales, $k\ll\mathcal{H}$, Weinberg's adiabatic theorem forbids the growth of the Newtonian potential \cite{Weinberg:2008zzc} in our single-field model. For a recent numerical study of the evolution of the curvature perturbation during preheating in multi-field models, see Ref. \cite{Jiang:2018uce}.} Therefore, while the inflaton condensate oscillates and we are in the linear regime for density perturbations, $\delta\rho/\bar{\rho}\ll1$, the growth in $\Phi$ induced by resonant instabilities does not lead to the breakdown of linear perturbation theory.

\subsubsection{Nonlinear regime}
\label{sec:NonLinReg}

The resonant growth in $\delta\phi$ eventually leads to the breakdown of linear perturbation theory, $\delta\rho/\bar{\rho}\sim1$. Backreaction comes from sub-horizon inflaton perturbations. The universe remains homogeneous and isotropic on super-horizon scales, whereas on sub-Hubble scales it becomes highly nonlinear. To estimate whether the self-gravity of sub-horizon nonlinear overdensities is important, we estimate the linear scalar metric perturbation they source in the homogeneous and isotropic spacetime, see eq. \eqref{eq:PertMetric}. To this end we consider a Poisson-like equation
\Beq
\label{eq:NonLinearPoisson}
\frac{\nabla^2 \Phi^{\rm nl}}{a^2}=\frac{\rho(x)-\bar{\rho}}{2\mpl^2}\,,
\Eeq
where the `nl' superscript reminds us that the gravitational sources are nonlinear. We ignore any possible general relativistic corrections, which normally is a good approximation on sub-horizon scales. Anisotropic nonlinear stresses, which lead to the distinction between two independent scalar metric perturbations (one of which is governed by eq. \eqref{eq:NonLinearPoisson}), are also not considered. Since the anisotropic stresses are comparable to $\rho$, we believe that $\Phi^{\rm nl}$ is sufficient for an order-of-magnitude estimate of the importance of the self-gravity of nonlinear sub-horizon structures.\\
\subsubsection{Oscillons \& Transients}
\noindent\textbf{\textit{Oscillons}: } To roughly estimate the scalar metric perturbations sourced by oscillons, we consider an individual object of mass $\mathcal{M}$ and radius $R$. Up to a constant, the Newtonian potential on the surface of a single oscillon is
 \Beq
\!\!|\Phi^{\text{nl}}|\sim \mathcal{M}/(8\pi\mpl^2 R)\sim\rho_{\text{c}} R^2/(6\mpl^2)\sim H_{\text{br}}^2R^2\,,
 \Eeq
 where $\rho_{\text{c}}$ is the core energy density and factors of order unity have been ignored. The last scaling needs an explanation. Oscillons formed as a consequence of self-resonance have $\phi_{\text{c}}\sim\bar{\phi}_{\text{br}}$, where $\bar{\phi}_{\text{br}}$ is the amplitude of the oscillating inflaton at the time of backreaction (or fragmentation). Hence, $\rho_{\text{c}}\sim\bar{\rho}_{\text{br}}$ and $\Phi^{\text{nl}}\sim H_{\text{br}}^2R^2$. In this very rough estimate, horizon size objects can become strongly gravitating. However, the typical scale of inhomogeneities at the time of backreaction tends to be much smaller than the Hubble horizon. The maximally resonant lengthscale $k_{\rm phys}^{-1}\sim R_{\text{res}}/(2\pi)\sim {\rm few}\, m^{-1}$, which we can take as an upper bound on the oscillon radius, i.e., $R\sim10m^{-1}$. At the time of backreaction, $H_{\text{br}}^2\sim m^2\bar{\phi}^2_{\text{br}}/(6\mpl^2)$. Putting these together
\Beq
\label{eq:PsiUpperBound}
 |\Phi^{\text{nl}}|\sim 10\times\left(\frac{M}{\mpl}\right)^2\,.
\Eeq
Since we require $M\ll \mpl$ for efficient resonance, it is unlikely for oscillons to have $\Phi^{\text{nl}}\gsim1$ and collapse due to their own gravitational field.

Besides the Newtonian potential, it is also useful to consider the gravitational field on the surface of the oscillons:
\Beq
\label{eq:gUpperBound}
g
&\equiv|\boldsymbol{\nabla}\Phi^{\rm nl}|/a\\
&\sim\Phi^{\text{nl}}/R\sim10(M/\mpl)^2/R\,.
\Eeq

\noindent\textbf{\textit{Transients}: }Similar considerations apply to transients, with $m=m(\phi=M)$. In particular, the predictions for the Newtonian potential  and acceleration on the surface of a single object given in eq. \eqref{eq:PsiUpperBound} and eq. \eqref{eq:gUpperBound}, respectively, still hold.

\begin{figure*}[t!] 
   \centering
\includegraphics[width=6.5in]{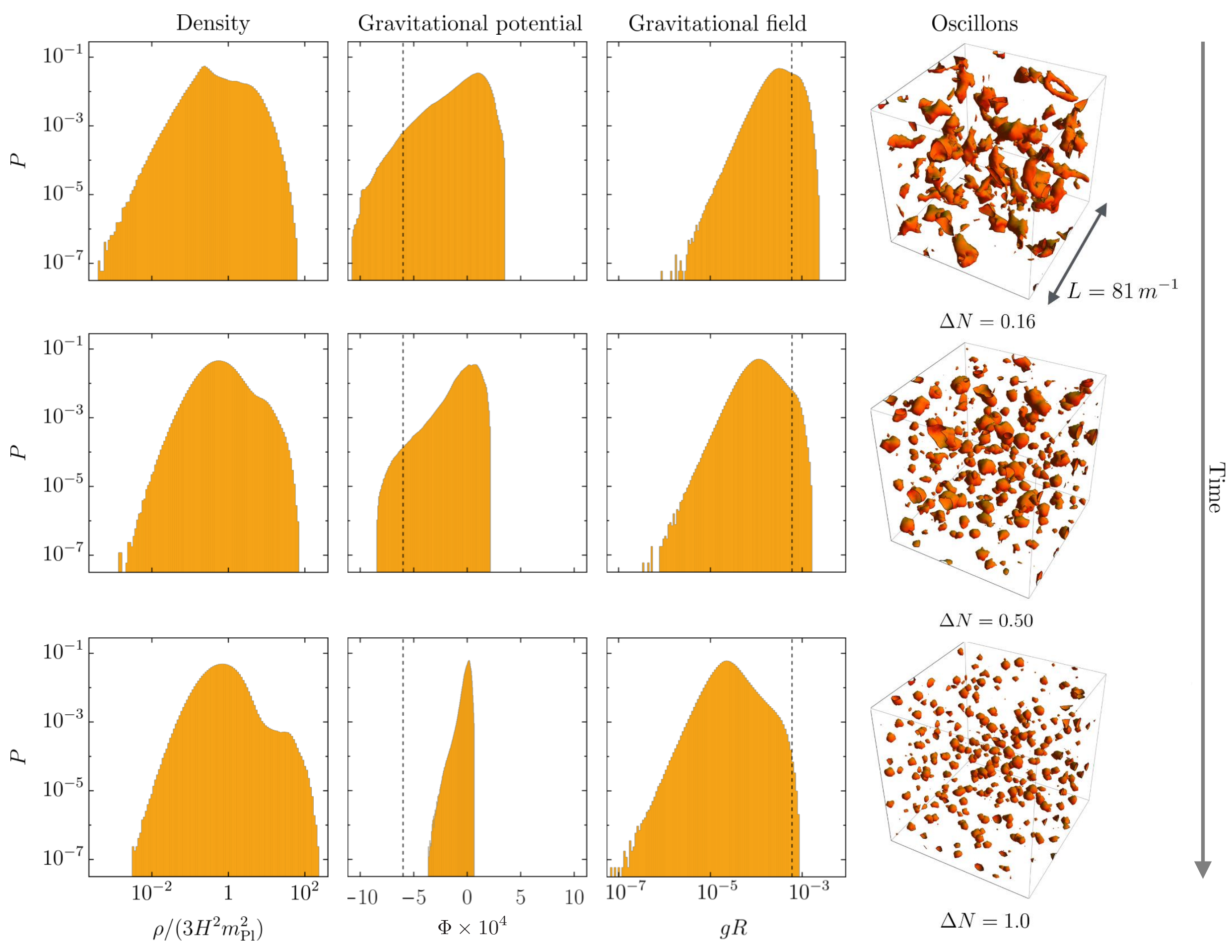}
   \caption{Formation of oscillons after inflation and their persistence. In each row we show the histograms of the energy density, $\rho$, Newtonian potential, $\Phi$, and the gravitational acceleration, $g$, across the simulation box at $\Delta N$ {\it e}-folds after the end of inflation (in each column, later times are at the bottom). The orange contours, in the snapshots of the simulation box in the last column, are drawn around regions of overdensity $\geq5$. This is for the T-model with $n=1$, $M=\sqrt{6\alpha}\mpl$, $\alpha=10^{-5}$. The vertical dashed line is at $gR=-\Phi=10(M/\mpl)^2$ -- the approximate prediction for the Newtonian potential on the oscillon surface of radius $R$. Since oscillons are spherical, localized objects, $g$ should be maximal near their surfaces. It agrees with the observed maximal value of $g$ within the simulation box.}
   \label{fig:OscHist}
\end{figure*}

\begin{figure*}[t!] 
\includegraphics[width=6.5in]{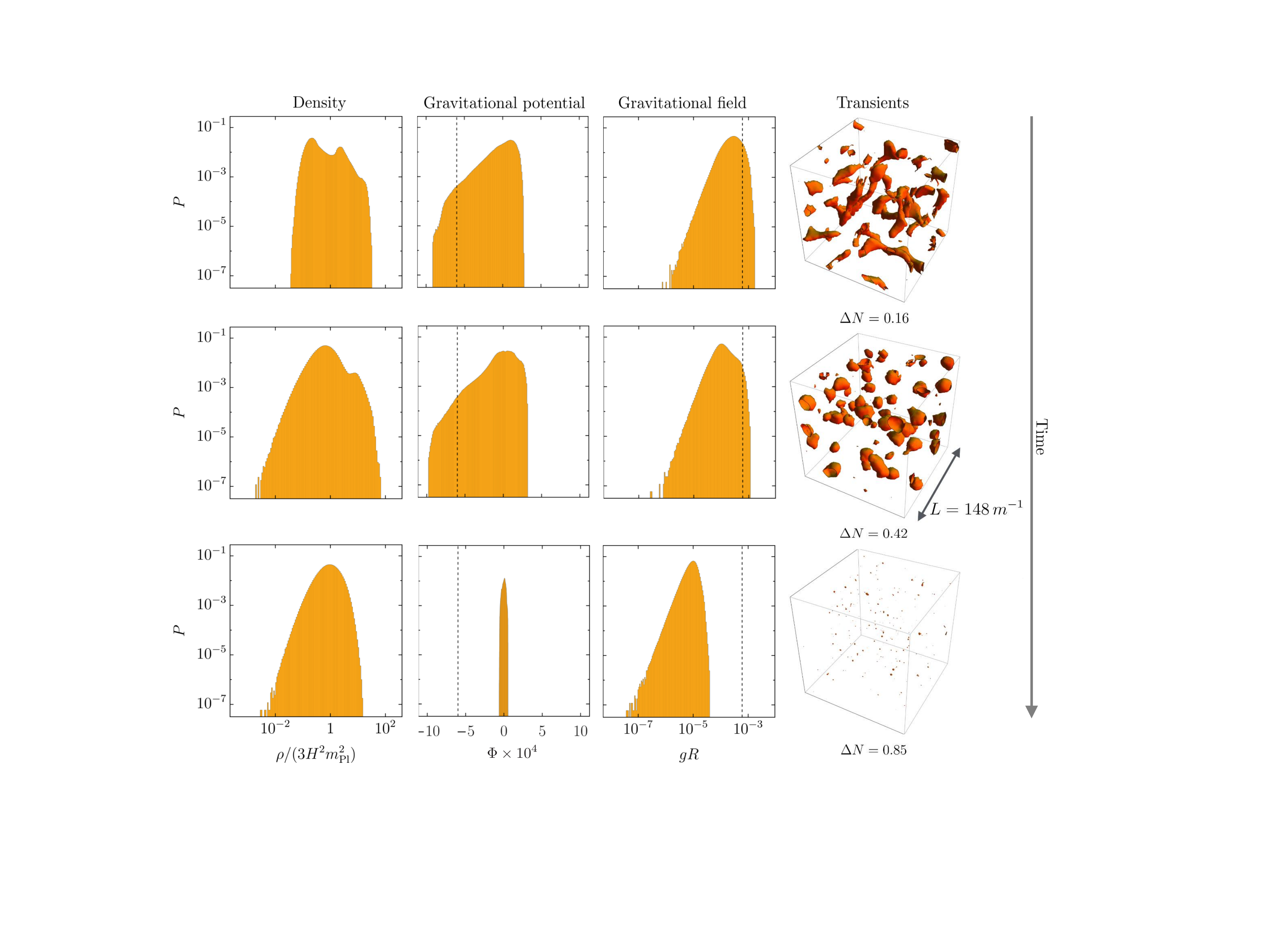}
   \caption{Transients formation and decay. The notation and parameters are the same as in Fig. \ref{fig:OscHist}, besides $n=2$.}
   \label{fig:TransHist}
\end{figure*}

\subsection{Results from lattice simulations}
\label{sec:GravScalLattSims}

We calculate the nonlinearly sourced scalar metric perturbation, $\Phi^{\rm nl}$ in eq. \eqref{eq:NonLinearPoisson},  by modifying {\it LatticeEasy } \cite{Felder:2000hq}. The inflaton field and the scale factor are evolved according to the prescription from the default version of {\it LatticeEasy } (with gravity only at the background level). The Newtonian potential is then calculated from the energy density of the fields using our own Poisson equation solver (described in Appendix \ref{sec:Poisson}). For our simulations, we consider parameter sets used in our previous studies \cite{Lozanov:2016hid,Lozanov:2017hjm} and repeat the same convergence tests. 
\subsubsection{Oscillons}
 In Fig. \ref{fig:OscHist}, we show histograms of the energy density, the Newtonian potential and the magnitude of the gravitational field across the box for $M\approx0.775\times10^{-2}\mpl$ at $0.16$, $0.50$ and $1.00$ {\it e}-folds after the end of inflation. The first row represents the moment when backreaction effects become important. The second one shows the beginning of oscillon formation and the third row shows the settled oscillon configuration. 

The histograms were generated after finding the greatest and the smallest values of the quantity of interest. Afterwards the interval was split into $200$ ($100$) linear (logarithmic) bins for the Newtonian potential (the energy density and $g$). The heights of the bins are normalized, i.e., $\sum\limits_{\rm bins}P=1$. 

As we discuss below, the numerical solution to eq. \eqref{eq:NonLinearPoisson} using data from our lattice simulations confirms the estimates from Section \ref{sec:NonLinReg}.\\ 

\noindent {\it Energy density}: \\
The evolution of the energy density (first column in Fig. \ref{fig:OscHist}) contains valuable information. In the top two panels of the first column, one can see the gradual formation of a high density plateau, which becomes very prominent in the bottom-most panel. This plateau in the histogram represents points lying within oscillons. Oscillon cores are highly overdense -- the maximal density seen in the second from the top panel is at $\rho_{\rm max}\approx60\times\bar{\rho}$, whereas in the bottom panel $\rho_{\rm max}\approx200\times\bar{\rho}$. Note that this increase in the oscillon overdensity is expected since the average energy density in the box redshifts as $\bar{\rho}\propto a^{-3}$, while the energy density inside oscillons is constant (i.e., does not redshift).

In the energy density histograms one can also see the formation of a well-defined low-density peak, centered near the mean density, $\bar{\rho}$. Unlike the high-density plateau, once formed it does not evolve with time. Most of it represents the underdense regions, lying outside the oscillons. It implies that oscillons, albeit dominant in energy, are subdominant in volume. The symmetry of this low density peak points towards a lognormal distribution of the energy density in the underdense regions which is a tantalizing hint for relativistic turbulence. Indeed, in the {\it Transients} subsection to follow, we show the development and persistency of a similar peak and its lognormality for transients. \\

\noindent{\it Newtonian Potential}:\\
The second column in Fig. \ref{fig:OscHist} shows the evolution of the Newtonian potential. Our naive estimate for the gravitational potential inside oscillons in eq. \eqref{eq:PsiUpperBound} (represented by a vertical dashed line in the second column panels) describes the typical value calculated from the simulation reasonably well (up to order unity).

The skewness of the Newtonian potential histograms is due to oscillons. Since oscillons represent highly overdense regions giving rise to gravitational attraction, we see greater departures of $\Phi^{\text{nl}}$ from $0$ in the negative direction. Note that we have put $\bar{\Phi}^{\text{nl}}=0$ in the simulation box. The small decrease in the greatest departure from $0$ between the second and third panels is because the oscillons in the second panel are slightly bigger and decay into smaller ones which form the stabilized configuration in the third panel. \\

\noindent{\it Gravitational field}:\\
In the third column of Fig. \ref{fig:OscHist}, we show the evolution of histograms of the gravitational field (equivalently, acceleration). If the oscillons had a uniform spherically symmetric density up to radius $R$, then $g\propto r$ for $r<R$ and $g\propto r^{-2}$ for $r>R$, where $r$ is the distance from the oscillon core. Hence, the maximal $g$ will be on the surface of the oscillons. Our oscillons do not have an exactly uniform density, but we still expect that the maximal $g$ in the histograms will come from regions close to the oscillon surfaces. This maximal value was estimated in eq. \eqref{eq:gUpperBound} and is represented by a vertical dashed line; it again agrees with the values from the numerical simulations.

Let us re-iterate the main takeaway from this subsection. Since oscillons do not form efficiently for $M\gtrsim10^{-2}\mpl$, the gravitational potential on the surfaces of individual objects is bound to be
\Beq
\label{eq:PsiSurface}
|\Phi^{\text{nl}}|\sim10\times\left(\frac{M}{\mpl}\right)^2\lesssim10^{-3}\,.
\Eeq
Oscillons do not gravitate strongly, justifying the linear treatment of metric perturbations. Nevertheless, it will be interesting to study the stability of these  individual objects and their clustering when the backreaction from gravity is included. Gravitational clustering could lead to collisions as well as collapse of overdense oscillon clusters and perhaps formation of PBHs \cite{Konoplich:1999qq,Khlopov:1999ys,Rubin:2000dq,Cotner:2018vug}, see Section \ref{sec:OscPBHs} for further discussion.

\begin{figure}[t] 
   \centering
   \hspace{-0.3in} 
   \includegraphics[width=3.0in]{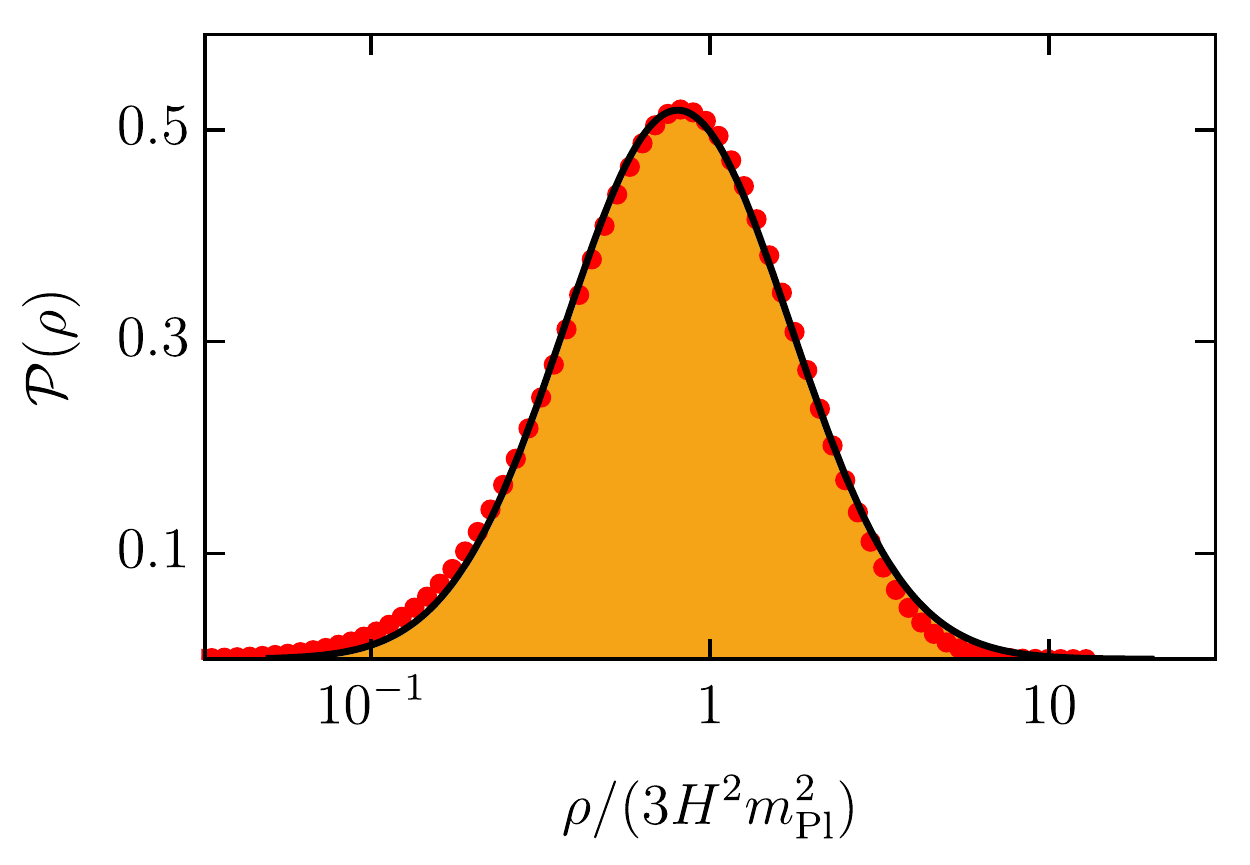}
   \caption{The energy density distribution for $n=2$, $M\approx0.775\times10^{-2}\mpl$, after the transients decay. The black curve and the orange-shaded area underneath represent a fit, see eq. \ref{eq:LogNormFit}, to the simulation data (red points).}
   \label{fig:HistTransients}
\end{figure}

\subsubsection{Transients}
\label{sec:Transients}
We have also analysed the scalar metric perturbations created by the transients, see Fig. \ref{fig:TransHist}. The data is for $n=2$, $M\approx0.775\times10^{-2}\mpl$, extracted at $0.16$, $0.42$ and $0.85$ {\it e}-folds after the end of inflation. The first row shows the moment when backreaction becomes significant. The second row depicts a relatively stable transient configuration and the third row shows the simulation box after the transients have disappeared. 

During the formation of the transients and their stabilization (first and second row) we arrive at qualitatively similar results to oscillons. We again observe the formation of a high-density plateau region and a broad low-density peak centered near $\bar{\rho}$, a skewed Newtonian potential distribution and maximal $|\Phi^{\text{nl}}|$ and $g$ in agreement with our expectations and similar to the oscillon case, see eq. \eqref{eq:PsiSurface}.

After the transients decay away (third row) we observe a very simple picture. The inflaton dynamics is dominated by the potential minimum, i.e., the evolution should be similar to the pure $\lambda\phi^4$ case studied in \cite{Khlebnikov:1996mc,Micha:2004bv,Micha:2002ey}. We find that the energy density histogram becomes quite symmetric. The high-density plateau disappears, whereas the broad peak centered near the mean remains. It can be fitted well with a lognormal curve, see Fig. \ref{fig:HistTransients}. There we fit the data from the first panel in the third row in Fig. \ref{fig:TransHist} with 
\Beq
\label{eq:LogNormFit}
P(\rho)&=\mathcal{P}(\rho)\Delta\ln\rho\,,\\
\mathcal{P}(\rho)&=\frac{1}{\sqrt{2\pi}\sigma}\exp\left[-\frac{(\ln[\rho/(3H^2\mpl^2)]-\mu)^2}{2\sigma^2}\right]\,,
\Eeq
where $\Delta\ln\rho$ is the width of the logarithmic bin. We also put $\mu=-\sigma^2/2$, so that the expectation value of $\rho$ is equal to the measured mean. This leaves only $\sigma$ as a free parameter and after fitting we find $\sigma\approx0.77$. It is remarkable that a one-parameter fit describes the data so well. The lognormality is a tantalizing hint for relativistic turbulence as pointed out in \cite{Frolov:2008hy,Frolov:2010sz}.

The Newtonian potential also becomes symmetric about $0$ after the transients decay (see the second panel in the third row in Fig. \ref{fig:TransHist}). This reflects the symmetric distribution of underdensities and overdensities about the mean. Note that the maximal departure from $0$ now is only $|\Phi^{\text{nl}}|\approx5\times10^{-4}$ which is an order of magnitude smaller than when transients (or oscillons) are present. The gravitational field, $g$, also decreases accordingly.

\section{Gravitational waves}
\label{sec:GWs}

The nonlinear, inhomogeneous field dynamics at the end of inflation sources gravitational waves. We first provide analytic estimates for the expected gravitational wave frequency and energy density from linear and nonlinear field dynamics. We compare these estimates with lattice simulations which compute the field dynamics and the gravitational wave production numerically. 
\subsection{Analytic estimates}
\subsubsection{Linear regime}

In the early post-inflationary stages, while the oscillating inflaton condensate is still intact, the scalar-vector-tensor decomposition applies. The linear tensor metric perturbations, $h_{ij}$, evolve freely. There are no constraint equations for them and they represent the two dynamical gravitational degrees of freedom. To linear order in metric perturbations, $h_{ij}$ is also gauge invariant, and is governed by the source-free linear equation of motion
\Beq
\label{eq:GWsHomEoM}
h_{ij}''+2\mathcal{H}h_{ij}'-\nabla^2 h_{ij}=0\,.
\Eeq
Just like during inflation, sub-horizon Fourier modes, $k\gg\mathcal{H}$, are oscillatory, with a decaying amplitude scaling inversely with the scale factor, $\propto a^{-1}$, whereas for super-horizon modes, $k\ll\mathcal{H}$, there is a constant (and a decaying) solution in accordance with Weinberg's adiabatic theorem \cite{Weinberg:2008zzc}. These GWs have a purely quantum origin. 

Earlier on, during inflation, co-moving modes lying deep within the Hubble sphere, started out in the Bunch-Davies vacuum, $\propto e^{-ik\tau}/(a\sqrt{k})$. Those which were stretched to super-Hubble scales froze (attaining a scale-invariant power spectrum), whereas those which did not cross out, remained in their vacuum state. 

Later on, during the oscillatory stage, the frozen modes remain constant until re-entry inside the horizon, where they start decaying and oscillating again (being in a non-vacuum state). Those in the Bunch-Davies vacuum remain in it. These modes are considered unphysical in the sense that their contribution to the energy budget of the universe is ignored. Otherwise, we enter the realm of the Cosmological Constant Problem.

\subsubsection{Nonlinear regime}

The fragmentation of the inflaton condensate due to resonant particle production can lead to the generation of a stochastic gravitational wave background \cite{Khlebnikov:1997di,Easther:2006vd,Easther:2006gt,GarciaBellido:2007af,Dufaux:2007pt,Dufaux:2008dn,Dufaux:2010cf,Adshead:2018doq}. The linear tensor metric perturbations are sourced by the nonlinear inflaton configurations
\Beq
\label{eq:GWsInhEoM}
h_{ij}''+2\mathcal{H}h_{ij}'-\nabla^2 h_{ij}=\frac{2}{\mpl^2}\Pi_{ij}^{TT}\,,
\Eeq
where the source term is the transverse traceless part of the anisotropic stress tensor
\Beq
\label{eq:AnisStr}
\Pi_{ij}^{TT}=\left(\partial_i\phi\partial_j\phi\right)^{TT}\,.
\Eeq
Note that unlike the gravitational waves from inflation, the GWs from the nonlinear stage have a classical origin. They are sourced by the classical evolution of inhomogeneities on sub-horizon scales. These GWs are the particular solution of the inhomogeneous equation of motion, eq. \eqref{eq:GWsInhEoM}, whereas the inflationary GWs are the complementary solution of the homogeneous part, eq. \eqref{eq:GWsHomEoM}. In this work we focus on the GWs from the nonlinear stage. As we will see, their power spectrum is strongly peaked around a single
frequency, determined by the fragmentation lengthscale. This fragmentation lengthscale corresponds to the wavelength of inflaton perturbations modes (typically sub-horizon) which first went nonlinear. \\

\noindent{\it Frequency of GWs}:\\
Let us start with a quick derivation of the frequency observed today of a GW signal with a co-moving wavenumber $k$, generated when the scale factor is $a_{\rm g}$. The frequency (in Hz) can be written as
\Beq
f_0&=\frac{1}{2\pi}\frac{k}{a_0}\\
 &=\frac{1}{2\pi}\frac{k}{a_{\rm{g}}\bar{\rho}_{\rm{g}}^{1/4}}\left(\frac{\bar{\rho}_{\rm{g}}}{\bar{\rho}_{\rm{th}}}\right)^{1/4}\frac{a_{\rm{g}}}{a_{\rm{th}}}\frac{a_{\rm{th}}}{a_{\rm{0}}}\bar{\rho}_{\rm{th}}^{1/4}\,,
\Eeq
where $a_{\rm th}$ and $\rho_{\rm th}$ are the scalefactor and energy density of the universe when the dominant components of the energy density are thermalized. We define a mean equation of state $\bar{\rho} a^{3(1+w)}=\rm{const}$ for $a_{\rm{g}}<a<a_{\rm{th}}$ and recall that entropy conservation for $a>a_{\rm{th}}$ implies
\Beq
\frac{\rho_{\rm{th}}^{1/4}a_{\rm{th}}}{\rho_{\rm{rel},0}^{1/4}a_{\rm{0}}}=\left(\frac{g_{\rm{th}}}{g_0}\right)^{-1/12}\,,
\Eeq
where $\bar{\rho}_{\rm{rel},0}$ is the energy stored in relativistic degrees of freedom today. $g_{\rm th}$ and $g_0$ are the relativistic degrees of freedom at $a_{\rm th}$ and today, respectively. All of this yields the well-known result \cite{Easther:2006gt,Dufaux:2007pt}
\Beq
\label{eq:f0GenDef}
f_0=\frac{1}{2\pi}\frac{k}{a_{\rm{g}}\bar{\rho}_{\rm{g}}^{1/4}}&\left(\frac{a_{\rm{g}}}{a_{\rm{th}}}\right)^{(1-3w)/4}\left(\frac{g_{\rm{th}}}{g_0}\right)^{-1/12}\\
    &\times\left(3\Omega_{\rm{rel},0}\right)^{1/4}\sqrt{\mpl H_0}\,.
\Eeq
It can be further simplified, by substituting for the known parameters $\Omega_{\rm{rel},0}=4.3\times10^{-5}h_{100}^{-2}$, $h_{100}=0.67$ \cite{Aghanim:2018eyx} and making the mild assumptions that $g_{\rm{th}}=10^2$ and $a_{\rm{g}}\approx a_{\rm{th}}$, to
\Beq
\label{eq:GWfreq0}
f_0=\frac{k}{a_{\rm{g}}}\frac{1}{\bar{\rho}_{\rm{g}}^{1/4}}\times4\times10^{10}\,\rm{Hz}\,.
\Eeq
{\it Frequency upper bound}:\\
We note that the frequency of the peak of the GW power spectrum, $f_{0,\rm max}$, has an upper bound. The peak is determined by the co-moving wavenumber, $k_{\rm max}$, for which the source term, eq. \eqref{eq:AnisStr}, has maximal power. The same wavenumber determines where the gradient energy density, $\rho_{{}_\nabla}$, has maximal power. Since during preheating parametric resonance amplifies Bunch-Davies vacuum fluctuations, we have $\bar{\rho}_{\rm{g}}^{1/4}\gtrsim \bar{\rho}_{{}_\nabla,{\rm g}}^{1/4}>k_{\rm max}/a_{\rm g}$. Hence
\Beq
\label{eq:f0UpperBound}
f_{0,\rm max}\lesssim 4\times10^{10}\,\rm{Hz}\,.
\Eeq
If the bound is saturated, the energy density of the unstable Bunch-Davies vacuum fluctuations is similar to the background energy density at the start of preheating. This renders invalid the semi-classical picture of quantum field fluctuations on top of a classical inflaton background and {\it quantum backreaction} has to be considered. Note that this result can be extended to any preheating scenario involving inhomogeneous (not necessarily scalar) classical field configurations, since for relativistic fields the anisotropic stress is always less than or comparable to the gradient energy.

\begin{figure}[t] 
   \centering
   \includegraphics[width=3in]{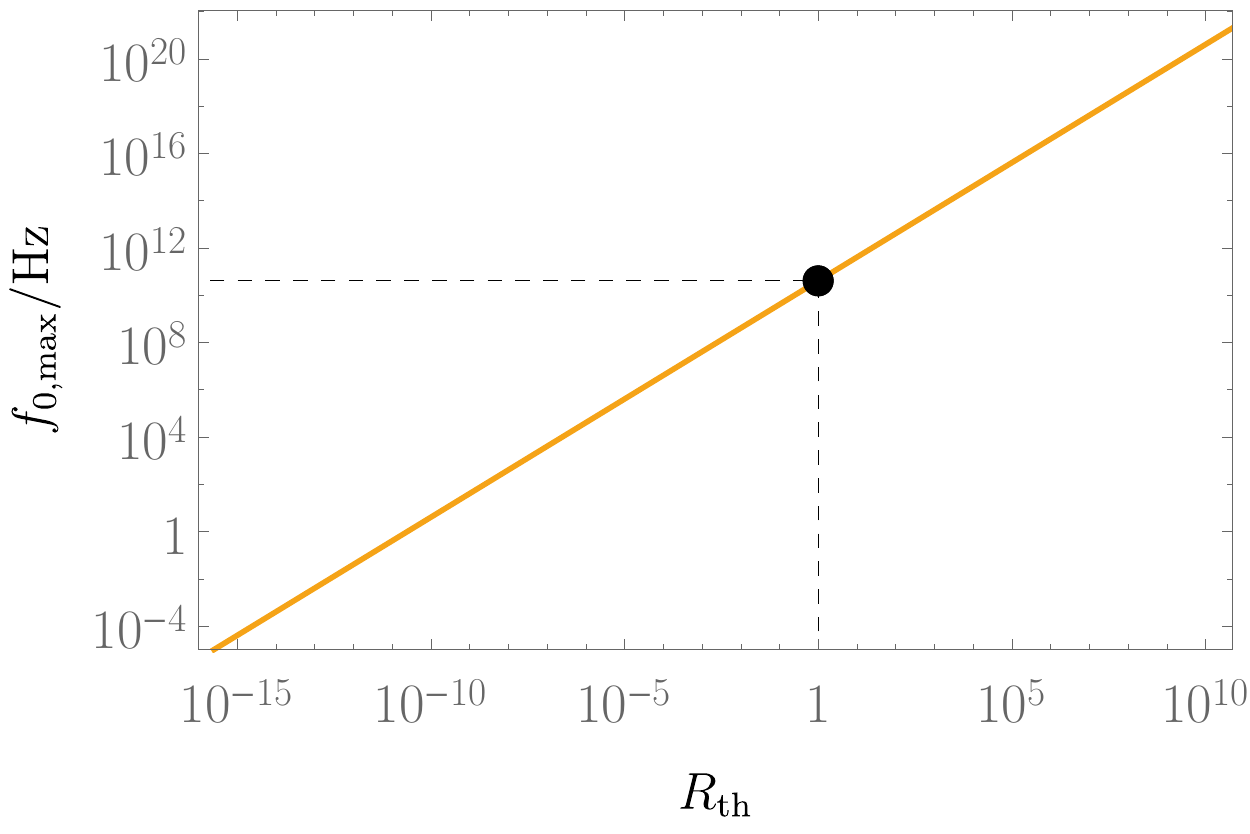}
   \caption{The upper bound on the today fequency of the maximally energetic GWs from preheating, $f_{0,\rm{max}}=4\times10^{10}R_{\rm{th}}\,\rm{Hz}$, $R_{\rm{th}}=(a_{\rm{g}}/a_{\rm{th}})^{(1-3w)/4}$, see eq. \eqref{eq:f0UpperBound} and preceding comments. In this paper we assume $R_{\rm{th}}=1$.}
   \label{fig:f0max}
\end{figure}

Thus, the maximal frequency at which we can reliably predict a gravitational wave signal from preheating is $\sim 10^{10}\,\rm{Hz}$. If we put aside all theoretical prejudice regarding the energy scale and dynamics at the time of preheating, then to completely explore the frequency range of gravitational waves possible from preheating, we need a detector with sufficient sensitivity up to $\sim 10^{10}\,{\rm Hz}$. 
\footnote{We stress that the frequency bound relies on the assumption of standard expansion histories after the generation of the GWs, i.e., $a_{\rm{g}}=a_{\rm{th}}$ or $w=1/3$ when $a_{\rm{g}}<a<a_{\rm{th}}$. If there is a period between the generation of GWs and thermalization with $w\neq1/3$, the frequency bound will receive a correction factor of $(a_{\rm{g}}/a_{\rm{th}})^{(1-3w)/4}$ according to eq. \eqref{eq:f0GenDef}, see Fig. \ref{fig:f0max}. In principle $g_{\rm{th}}$ should also be considered as a free parameter. However, since it enters with a power of $1/12$, only very exotic scenarios with a huge number of degrees of freedom can lead to correction factors significantly different from $1$. That is why we always keep $g_{\rm{th}}=10^2$.}

By looking at eq. \eqref{eq:GWfreq0} one can see two competing effects. For the peak frequency, the first factor giving the physical wavenumber at the time of the generation of the GWs is typically proportional to the Hubble parameter at that moment
\Beq
\frac{k}{a_{\rm{g}}}\equiv\beta^{-1}H_{\rm{g}}\,.
\Eeq
The earlier the GWs are generated, the smaller the horizon at that time is and hence the higher the frequency is. 

The energy density in the second factor ($\bar{\rho}_{\rm g}^{-1/4}$) has the opposite effect. The more efficient the resonance is (i.e., the earlier the time of backreaction is) the greater the mean energy density of the universe is and hence the lower the frequency is. This second effect is a manifestation of the redshifting of the gravitational waves -- the earlier in time they are generated the longer they are redshifted to lower frequencies. 

Overall, the first effect wins, since the Hubble parameter is proportional to the square root of the mean energy density. Hence, if we wish to drive $f_{0,\rm max}$ coming from eq. \eqref{eq:GWfreq0} to small enough values to be of observational interest for the model under consideration, we need to consider `inefficient' self-resonance (i.e., slow particle production, leading to delay in backreaction and late-time production of GWs). This can be seen explicitly in the following expression
\Beq
\label{eq:GWfreqbeta}
f_{0,\rm max}=\beta^{-1}\sqrt{\frac{H_{\rm{g}}}{\sqrt{3}\mpl}}\times4\times10^{10}\,\rm{Hz}\,.
\Eeq
The typical value of $\beta$ for self-resonance is $10^{-2}$, hence an energy scale of generation $\sim10^3\,\rm{TeV}$ (much lower than the typical large scale inflation) leads to $f_{0,\rm{max}}\sim1\,\rm{Hz}$. \\

\noindent{\it GW energy density}:\\
We now consider the actual strength of the GW signal itself. Conventionally, it is characterized by the ratio of the GW energy density by logarithmic co-moving momentum interval and the critical energy density of the universe, today, i.e.,
\Beq
\Omega_{\rm{GW},0}h_{100}^2=\frac{h_{100}^2}{\bar{\rho}_{\rm{c},0}}\frac{d \rho_{\rm{GW},0}}{d\ln k}\,.
\Eeq
We are again interested in the value of this quantity for the peak of the GW background. Since $\rho_{\rm{GW}}a^4=\rm{const}$, we get
\Beq
\label{eq:OmegaGW}
\Omega_{\rm{GW},0}h_{100}^2=\Omega_{\rm{GW,g}}&\left(\frac{a_{\rm{g}}}{a_{\rm{th}}}\right)^{1-3w}\\
                                                                                             \times&\left(\frac{g_{\rm{th}}}{g_{0}}\right)^{-1/3}\Omega_{\rm{rel,0}}h_{100}^2\,.
\Eeq
where $\Omega_{\rm GW,g}$ is the fractional energy density in gravitational waves at the time of production. We now estimate this energy density as follows:
\Beq
\rho_{\rm{GW,g}}\sim\left(\frac{h_{ij}'}{a_{\rm{g}}}\right)^2\mpl^2\sim\left(\frac{\nabla h_{ij}}{a_{\rm{g}}}\right)^2\mpl^2\,,
\Eeq
and from the equation of motion for gravitational waves, eq. \eqref{eq:GWsInhEoM},
\Beq
\frac{h_{ij}''}{a_{\rm{g}}^2}\sim\frac{\nabla^2 h_{ij}}{a_{\rm{g}}^2}\sim \frac{(\partial_i\phi\partial_j\phi)^{TT}}{\mpl^2}\,.
\Eeq
If a fraction $\delta_{{}_{\nabla}}$ of the mean energy density at the time of GW generation is stored in the form of gradients (taken as a proxy for the energy density involved in generating gravitational waves) then
\Beq
\label{eq:rhoGW}
\rho_{\rm{GW,g}}\sim H_{\rm{g}}^2\mpl^2\left(\frac{H_{\rm{g}}}{k/a_{\rm{g}}}\right)^2\delta_{{}_{\nabla}}^2\,\,\,
                               \sim\,\,\,\bar{\rho}_{\rm{c,g}}\beta^2\delta_{{}_{\nabla}}^2\,,
\Eeq
implying $\Omega_{\rm{GW,g}}\sim\beta^2\delta_{{}_{\nabla}}^2$. After substituting for it in eq. \eqref{eq:OmegaGW} we arrive at
\Beq
\label{eq:Ogw1}
\Omega_{\rm{GW},0}h_{100}^2\sim\beta^2\delta_{{}_{\nabla}}^2&\left(\frac{a_{\rm{g}}}{a_{\rm{th}}}\right)^{1-3w}\\
                                                                                             \times&\left(\frac{g_{\rm{th}}}{g_{0}}\right)^{-1/3}\Omega_{\rm{rel,0}}h_{100}^2\,.
\Eeq
Plugging in the values of the known parameters and setting $a_{\rm{g}}\approx a_{\rm{th}}$, finally yields
\Beq
\label{eq:OmegaGWfinal}
\Omega_{\rm{GW},0}h_{100}^2\sim10^{-5}\beta^2\delta_{{}_{\nabla}}^2\,.
\Eeq
In the calculations below, we will use $\delta_{{}_{\nabla}}=1/3$ for the inhomogeneous scalar field, typically valid at the time of backreaction of the field.

The typical values of $\Omega_{\rm{GW},0}h_{100}^2\lesssim10^{-10}$ at the peak are quite small. Qualitatively, this bound can be understood from the following reasoning. The factor of $10^{-5}$ in eq. \eqref{eq:OmegaGWfinal} comes from $\Omega_{\rm{rel,0}}$. Since gravitational waves redshift as radiation (or relativistic matter) we expect $\Omega_{\rm{GW}}$ to scale linearly with $\Omega_{\rm{rel}}$, which has been decreasing since the epoch of equality. The additional $\beta$ suppression is a consequence of the suppression of GW production on subhorizon scales sourced by the anisotropic part of the energy momentum tensor of the scalar field (see eq. \eqref{eq:rhoGW}). This last suppression is similar in nature to the one discussed after eq. \eqref{eq:GenPoisson} for the scalar metric perturbations. \\
\subsubsection{Oscillons \& Transients}
\noindent \textbf{\textit{Oscillons}: }For the typical lengthscale which first becomes nonlinear when oscillons form, the parameter $\beta$ is given by (refer to Section \ref{sec:NonLinReg})
\Beq
\beta=\frac{H_{\rm br}a_{\rm br}}{k}\sim\frac{H_{\rm br}R}{2\pi}\sim \frac{M}{\mpl}\,.
\Eeq
Assuming that the peak of the GWs is generated around the time of backreaction of this mode, its frequency today is
\Beq
\label{eq:fOscillons}
f_{0}\sim \sqrt{\frac{\mpl}{M}}\times 10^{8}\,\rm{Hz}\,.
\Eeq
In deriving the above expression we used eq. \eqref{eq:GWfreqbeta} and $H_{\rm br}\sim \Lambda^2/m_{\rm pl}$ with $\Lambda^2$ given by eq. \eqref{eq:Lambda4}. 

Similarly, using eq. \eqref{eq:OmegaGWfinal}, the expected strength of the gravitational waves today is
\Beq
\label{eq:OmegaOscillons}
\Omega_{{\rm GW},0}h^2_{100}\sim 10^{-6}\left(\frac{M}{\mpl}\right)^2\,.
\Eeq
Once oscillons have settled, we do not expect significant emission of GWs from individual oscillons, since field profiles of individual objects are spherically symmetric \cite{Amin:2018xfe}. 

We stress that if the universe is not radiation dominated after the time of production (which is likely since oscillons lead to a matter-like equation of state), then there will be additional suppression factors in the frequency (see eq. \eqref{eq:f0GenDef}) and the fractional density of the gravitational waves (see eq. \eqref{eq:Ogw1}) from oscillons after inflation. \\

\noindent \textbf{\textit{Transients}: }The formation of transients is very similar to the one of oscillons. We expect the frequency and the strength of the peak of the GW power spectrum to be the same as in the oscillon case, see eqs. (\ref{eq:fOscillons},\ref{eq:OmegaOscillons}). As the transients decay, those which evolve in a non-spherical manner may generate an additional GW signal. Its typical frequency should be again set by the spatial extend of the individual objects, whereas its strength is hard to model analytically and is best studied numerically. 
\subsection{Results from lattice simulations}
\label{sec:GWsOsc}
We employed {\it HLattice} \cite{Huang:2011gf} for the calculation of the GWs sourced by the nonlinear field dynamics. For the cases we studied, we used the same simulation parameters, i.e., box size, lattice points separation, time step, initial conditions, etc., as for the {\it LatticeEasy} simulations discussed in Section \ref{sec:GravScalLattSims}. However, we used a more accurate $6^{\rm th}$-order symplectic integrator for the self-consistent evolution of the scalar field and the scale factor. We also used the {\it HLattice2} spatial-discretization scheme (with $k_{\rm eff}$, not $k_{\rm std}$) when calculating the field spatial derivatives. Those improvements in accuracy were necessary for the computation of the GWs. To find the GWs, we evolved the tensor metric perturbations passively, i.e., we solved eq. \eqref{eq:GWsInhEoM}, without taking into account their feedback on the field and background metric dynamics. The time step for the GW integrator was four times greater than the one for the field and scale factor evolution. \\

\begin{figure}[t] 
   \centering
   \includegraphics[width=3.0in]{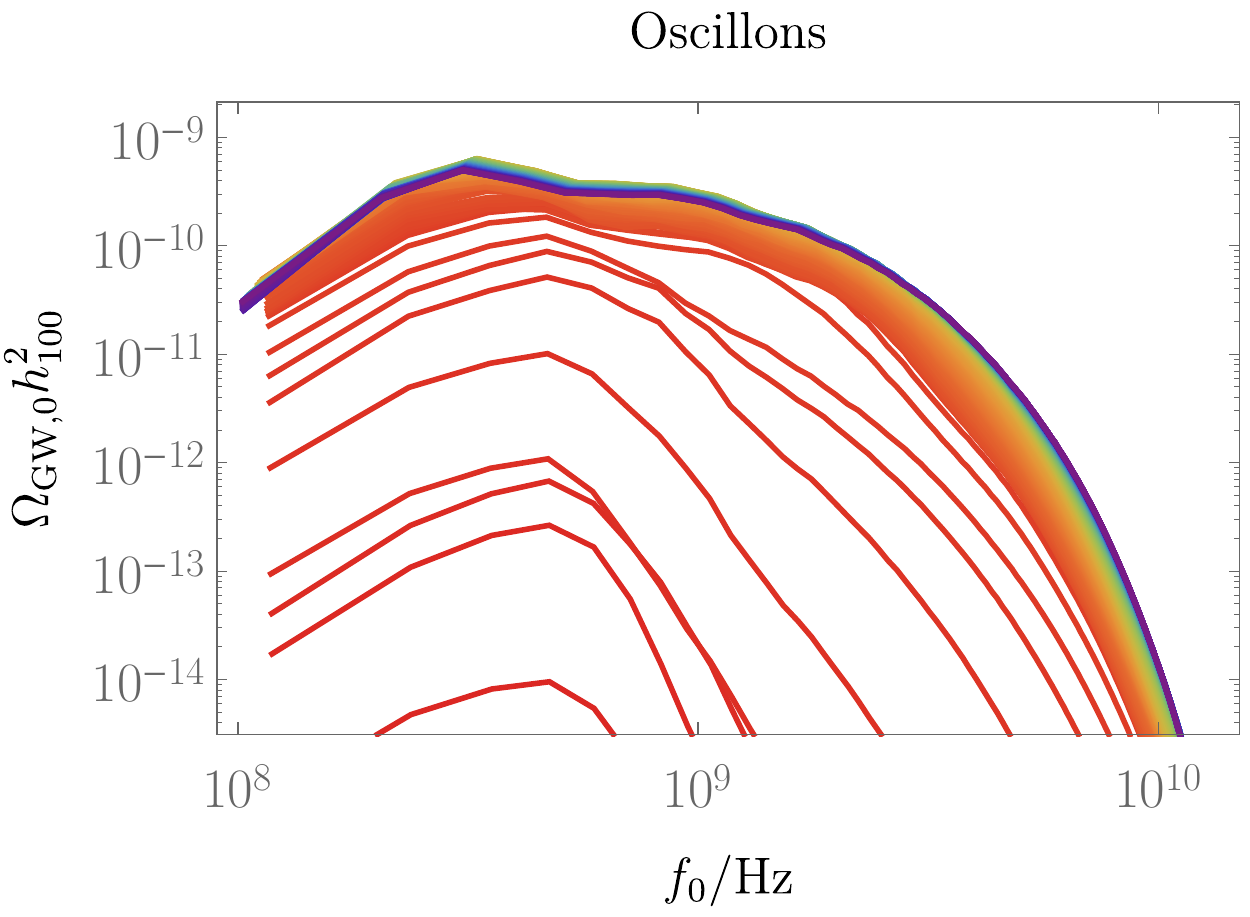}
   \caption{The gravitational waves generated between $\Delta N=0$ to $1$ (red to purple curves) for the oscillon model from Fig. \ref{fig:OscHist}. The peak of the red curves is close to the predicted values in eqs. \eqref{eq:fOscillons} and \eqref{eq:OmegaOscillons}.}
   \label{fig:GWsOscillons}
\end{figure}
\subsubsection{Oscillons}
The generated GW power spectrum from the oscillon formation for $M\approx0.775\times10^{-2}\mpl$ is shown in Fig. \ref{fig:GWsOscillons}. Time runs from red to purple. One can see four distinct stages \cite{Zhou:2013tsa}. 

The first $5-6$ red peaked curves represent the oscillatory stage, during which the condensate is still intact. A broad range of $\delta\tilde{\phi}_{\bf k}$ is steadily excited via broad resonance, see Section \ref{sec:ResPartProd}, and is responsible for the generation of the GWs. The frequency of the curves peak is slightly under $10^{9}\,\rm{Hz}$, which corresponds to the predicted order of magnitude in eq. \eqref{eq:fOscillons} and is determined by the wavenumber of the most unstable $\delta\tilde{\phi}_{\bf k}$. The rapid growth of the peak height reflects the exponential amplification of the inflaton perturbations. Even at this stage, the source term in eq. \eqref{eq:GWsInhEoM} has to be evaluated beyond linear order in perturbations.

The next $3-4$ red curves show the onset of the nonlinear regime. This stage is known as rescattering, since mode-mode couplings, including the backreaction of amplified $\delta\tilde{\phi}_{\bf k}$ on the condensate, become important. The broad peak, centered on the most unstable frequency, becomes wider. Its height grows more slowly than before and approaches the predicted value of $\sim10^{-10}$, see eq. \eqref{eq:OmegaOscillons}, as the field becomes completely inhomogeneous (with $\sim1/3$ of the total energy being stored in gradients). 

The following thick band of red-green curves represents the third stage. There the oscillons form and stabilize, with GWs power increasing slowly on all scales.

The last and longest stage is given by the green-purple curves. The oscillons have stabilized and sphericalized, while being assembled in a fixed co-moving grid-like configuration. Since there are almost no time-dependent quadrupole moments to act as sources, there is very little and slow production of GWs. On intermediate and low frequencies, GW power propagates (almost freely) towards lower frequencies and lower values as time goes by and the universe expands. This makes sense since the oscillon-dominated universe undergoes a matter-like state of expansion, with $\bar{\rho}\propto a^{-3}$. Since {\it HLattice} uses a formula like eq. \eqref{eq:GWfreq0} to calculate the GW frequency today (more specifically, $f_0(k,\tau)=k/(a(\tau)\bar{\rho}^{1/4}(\tau))\times4\times10^{10}\,\rm{Hz}$, where $\tau$ is the time of output, beyond which it is assumed that the universe is thermal and radiation dominated), it follows that $f_0(k,\tau)$ will decrease with time in a matter-dominated universe. The energy density of GWs redshifts as radiation, which explains why the GWs contribution to the energy budget of the matter-dominated universe decreases with time. Albeit nearly-spherical, individual oscillons do generate small amounts of GWs. This is visible at the high frequency end of the GW spectrum. Oscillons act as objects of fixed physical size, sourcing GWs of fixed physical wavenumber. For the {\it HLattice} conventions this implies that $f_0(k,\tau)\bar{\rho}^{1/4}(\tau)\propto k/a(\tau)=\rm{const}$, i.e., the oscillons-sourced GWs are at increasingly higher $f_0(k,\tau)$. This small late-time effect has an intrinsic numerical component. The oscillons are inevitably less well resolved as the comoving lattice expands, sourcing weak late-time high-frequency GWs. This does not affect the spectrum on intermediate and low frequencies. For more detailed studies of GWs from oscillons see \cite{Zhou:2013tsa,Antusch:2016con,Antusch:2017flz,Antusch:2017vga,Amin:2018xfe,Kitajima:2018zco,Liu:2018rrt}.\\
\begin{figure}[t] 
   \centering
   \includegraphics[width=3.0in]{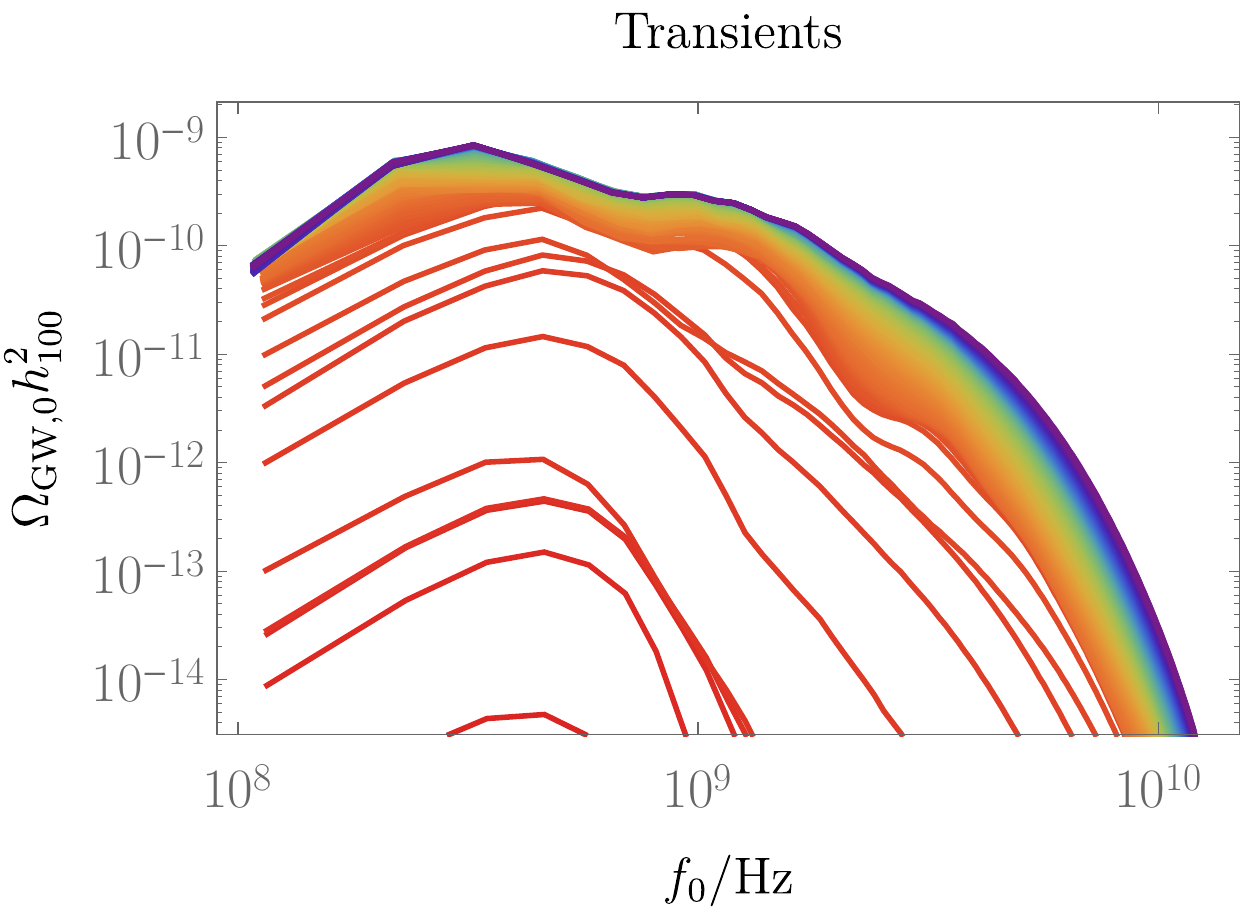}
   \caption{The gravitational waves generated between $\Delta N=0$ to $0.85$ (red to purple curves) for the transients model from Fig. \ref{fig:TransHist}. The peak of the red curves is close to the predicted values in eqs. \eqref{eq:fOscillons} and \eqref{eq:OmegaOscillons}, and almost identical to the one in Fig. \ref{fig:GWsOscillons}.}
   \label{fig:GWsTransients}
\end{figure}
\subsubsection{Transients}
Transients decay away quickly, in a non-spherical manner. Hence, unlike the cases when we have oscillons in which gravitational waves are not generated after oscillons are formed, the decay of the transients potentially can act as an additional source of GWs. Let us see what the simulations actually tell us. In Fig. \ref{fig:GWsTransients} we give the evolution of the GWs power spectrum for $M\approx0.775\times10^{-2}\mpl$, $n=2$ (the notation is the same as in Fig. \ref{fig:GWsOscillons}). One can distinguish five different stages. The first three are identical to the oscillon case: steady generation of GWs due to self-resonance (peaked around the frequency from eq. \eqref{eq:fOscillons}), a stage of rescattering when backreaction effects become important (with the peak widening and approaching the value from eq. \eqref{eq:OmegaOscillons}), and a stage where the transients form and stabilize and GWs are produced slowly on all frequencies. However, the fourth stage (of transients decay) and the fifth stage (of turbulent evolution) are not observed with oscillons.

While transients are decaying (green lines in Fig. \ref{fig:GWsTransients}), GWs are generated on high frequencies and within a broad low-frequency band. Note that the equation of state at these times is still $<1/3$ \cite{Lozanov:2016hid,Lozanov:2017hjm} and the GWs energy density red-shifts faster than $\bar{\rho}$. This additional sourcing of GWs provides some boost of the signal (by a factor of a few) in comparison to the oscillon case.

After the transients decay, their dynamics is dominated by the potential minimum. We find the same type of GWs (light blue-purple lines in Fig. \ref{fig:GWsTransients}) as observed in the pure $\lambda\phi^4$ case \cite{Khlebnikov:1997di,Easther:2006gt,GarciaBellido:2007af}. There is no generation of GWs on lower frequencies, whereas there is some continuous production at higher frequencies.

\section{Discussion}
\label{sec:Disc}

\subsection{PBH seeds}
\label{sec:OscPBHs}

The sourced gravitational field by oscillons is weak. The magnitude of the Newtonian potential on their surfaces is given in eq. \eqref{eq:PsiUpperBound}. It cannot exceed $10^{-3}$, since oscillons cannot form for $M\gtrsim10^{-2}\mpl$, see Refs. \cite{Lozanov:2016hid,Lozanov:2017hjm}. Strong gravitational effects such as gravitationally induced collapse of individual objects, which may lead to the formation of primordial black holes (PBHs), is expected only for large values of the scalar potential, $|\Phi|\sim1$. Since this is not the case, the self-gravity of individual objects affects their shapes negligibly.

However, the gravitational potential of an  overdense region of a collection of oscillons can potentially grow over time. This longterm effect (which lies well beyond the duration of our simulations) can potentially lead to the formation of PBHs \cite{Cotner:2018vug}. Overdense clusters of oscillons can in principle collapse due to their own gravity, however the details would depend on angular momentum distribution as well as the details of close interactions. Some of the collapsed cluster objects can become PBHs which can account for a significant fraction of the dark matter in our universe and/or the binary black hole merger events observed in Ligo \cite{Abbott:2016blz,TheLIGOScientific:2016wyq,Bird:2016dcv,Clesse:2016vqa,Cotner:2018vug}. We note that this growth of overdensities, at least before close encounters of the solitons, is similar in nature to the one of a massive oscillating and self-gravitating inflaton condensate \cite{Easther:2010mr}.

Unlike oscillons, we do not expect transients to act as efficient seeds of PBHs for two reasons. Firstly, transients form when the scalar field potential near the minimum $\propto|\phi|^{2n}$, $n>1$, implying positive pressure of the effective background fluid, $p>0$ \cite{Lozanov:2016hid,Lozanov:2017hjm}. This can lead to a substantial suppression of the growth rate of overdensities \cite{Polnarev:1986bi} (note for oscillons $p\approx0$ \cite{Lozanov:2016hid,Lozanov:2017hjm}). Secondly and more importantly, the transients are quite unstable. It is unlikely that they can be present for the entire duration of the slow collapse of a cluster of objects. Hence, the PBHs formation mechanism employed for oscillons simply does not apply to transients, due to the short livetimes and positive background pressure of the latter. 

Nevertheless, one may wonder whether the inherent instability of the transients (which leads to their collapse due to their self-interactions) provides another way for forming PBHs. We have checked that during the collapse of individual objects, the gravitational potential does not change much (i.e., its magnitude never exceeds $10^{-3}$ for the optimal choice of parameters). Just like oscillons, individual transients cannot collapse into PBHs.

\subsection{Gravitational wave detection}
\label{sec:ObsGWs}

\begin{figure}[t] 
   \centering
   \includegraphics[width=3.1in]{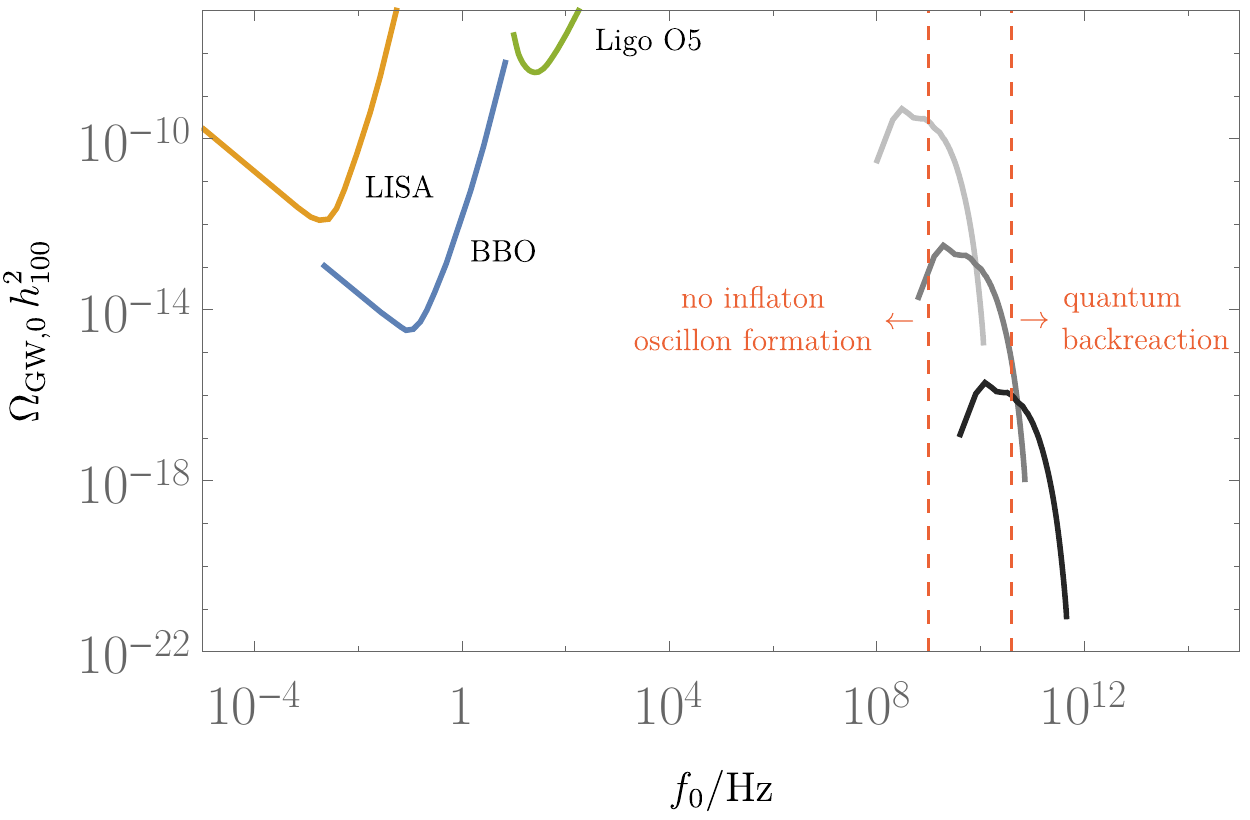}
   \caption{The gray, dark gray and black lines show the variation of the oscillon-sourced GWs power spectra today, scaled according to  eqs. \eqref{eq:fOscillons} and \eqref{eq:OmegaOscillons}. For $M$ yielding peak frequencies lower than the one of the left red dashed line, the inflaton does not form oscillons. For $M$ leading to frequencies greater than the one of the right red dashed line the consistency of oscillon formation scenario is ruled out by virtue of the contraint from eq. \eqref{eq:f0UpperBound}.  The detection ranges of LISA, BBO and Ligo O5 \cite{Moore:2014lga,TheLIGOScientific:2016wyq} are given in orange, blue and green, respectively.}
   \label{fig:GWsObs}
\end{figure}

The formation of oscillons leads to the generation of a stochastic gravitational wave background. In Fig. \ref{fig:GWsObs} we give its power spectrum for three different values of $M$ in gray, dark gray and black. The gray curve is from the end of the simulation presented in Section \ref{sec:GWsOsc} (for $M\approx0.775\times10^{-2}\,\mpl$). The dark gray and black curves are for $M_1\approx2.44\times10^{-4}\mpl$ and $M_2\approx6.25\times10^{-6}\mpl$, respectively. They were found after rescaling the gray curve according to $f_0\rightarrow(f_{0,\rm{max}}(M_{1,2})/f_{0,\rm{max}}(M))f_0$ and $\Omega_{\rm{GW},0}\rightarrow(f_{0,\rm{max}}(M)/f_{0,\rm{max}}(M_{1,2}))^4\Omega_{\rm{GW},0}$. The first rescaling is trivial, whereas the second one comes from eqs. \eqref{eq:fOscillons} and \eqref{eq:OmegaOscillons}. 

Since the inflaton does not form oscillons when $M\gtrsim10^{-2}\mpl$, see Refs. \cite{Lozanov:2016hid,Lozanov:2017hjm}, the gray curve is close to the lowest in frequency (and the greatest in strength) signal that can be achieved within the allowed parameter space. It has $f_{0,\rm{max}}\sim10^{9}\,\rm{Hz}$ as depicted by the left red dashed vertical line in Fig. \ref{fig:GWsObs}. The right red dashed vertical line corresponds to the upper bound given in eq. \eqref{eq:f0UpperBound}. It is saturated by $M_2$, see eq. \eqref{eq:fOscillons}.

There is a very narrow frequency range in which we can detect GWs from inflaton oscillons. It lies well beyond the reach of planned GW detectors, see Fig. \ref{fig:GWsObs}. However, the conclusions drawn so far rely on two major assumptions: (i) the universe thermalises (and becomes radiation dominated) immediately after the generation of GWs, $a_{\rm g}=a_{\rm th}$, see eqs. \eqref{eq:f0GenDef} and \eqref{eq:OmegaGW}; (ii) the inflaton, not another spectator scalar field, forms the oscillons. 

If we relax (i) (but keep (ii)) and put $a_{\rm g}<a_{\rm th}$, as well as $w<1/3$ in eqs. \eqref{eq:f0GenDef} and \eqref{eq:OmegaGW}, the GWs signal can be brought to frequencies of observational interest, see Figs. \ref{fig:f0max} and \ref{fig:GWsObs}. Nevertheless, the strength of the signal will be below the sensitivity of the planned detectors. On the other hand, if we relax (ii), while keeping (i), the frequency and the strength of the signal will be determined by the generic expressions given in eqs. \eqref{eq:GWfreqbeta} and \eqref{eq:OmegaGWfinal} (not in eqs. \eqref{eq:fOscillons} and \eqref{eq:OmegaOscillons}). This way the frequency of the GW signal can be brought down to sufficiently low values, without penalizing the strength of the signal, making planned GW detectors the ideal probes of oscillons \cite{Antusch:2016con,Antusch:2017flz,Antusch:2017vga,Amin:2018xfe,Kitajima:2018zco,Liu:2018rrt}.

Even if the oscillon-sourced GWs elude direct detection, in principle they can still have observational consequences. Since GWs are massless, they contribute to the effective number of light degrees of freedom, $\Delta N_{\rm eff}$, during the epoch of the CMB. Thereby, they can affect the Big Bang Nucleosynthesis and the cosmic microwave background \cite{Maggiore:1999vm}. More specifically, bounds on Beyond-the-Standard-Model contributions to $\Delta N_{\rm eff}$ constrain the total integrated GW energy density. If primordial GWs are the only relativistic degrees of freedom apart from the Standard Model ones, present constraints yield $\int{\rm d}\ln f\, \Omega_{{\rm GW},0}(f)\lesssim 10^{-6}$ \cite{Pagano:2015hma}. The upcoming CMB S4 experiments are expected to improve the threshold by over an order of magnitude \cite{Abazajian:2016yjj}. The oscillon-sourced GWs $\int{\rm d}\ln f\, \Omega_{{\rm GW},0}(f)\sim 10^{-9}$, require an additional couple of orders of magnitude to become detectable.\footnote{On the one hand, this is difficult because it is likely that there is an interval of matter domination following the time when gravitational wave from oscillons are produced (though not true for transients). Alternatively, one can assume an expansion period with a stiff equation of state, $w>1/3$, \cite{Figueroa:2018twl} between the oscillon decay and thermalization, to boost $\Omega_{{\rm GW},0}$.} Note that this measurement is insensitive to the peak frequency and provides a unique window to high-frequency GWs, which are otherwise inaccessible with current and planned GW detectors. 

Despite transients being unstable and decaying away the same conclusions, regarding the observations of the stochastic gravitational wave background they generate, apply to them.

\section{Conclusions}
\label{sec:Conclusions}

We studied the scalar and tensor metric perturbations generated by oscillons and transients after inflation. We evolved the nonlinear inflaton structures numerically with classical lattice simulations. We used the simulation output to source the metric perturbations passively, i.e., without considering the backreaction of the metric perturbations on the inflaton dynamics. However, gravity was included in the lattice simulations at the background level.

For the magnitude of the scalar metric perturbations we found a parametrized upper bound, see eq. \eqref{eq:PsiSurface}. It is the same for transients and oscillons. For the optimal choice of parameters, it is $\approx10^{-3}$, implying that both oscillons and transients are weakly gravitating objects. It is highly unlikely for individual objects to collapse under their own gravitational pull and form PBHs.

The spectra of the GWs sourced by oscillons and transients are quite similar. Albeit the transients decay generates additional GWs and the subsequent evolution of the turbulent inflaton leads to even more GWs, the shapes, the height and the frequency of the sourced GW power spectrum by oscillons and transients are almost the same, see Figs. \ref{fig:GWsOscillons} and \ref{fig:GWsTransients} and eqs. \eqref{eq:fOscillons} and \eqref{eq:OmegaOscillons}. If the oscillons or the transients are formed by the inflaton, and we assume that the universe becomes radiation dominated soon after soliton formation, the typical GWs frequencies today are $10^{9}-10^{10}\,\rm{Hz}$. They lie well beyond the reach of all planned GW detectors. If we postulate a long period of $w<1/3$ state of expansion after the inflaton fragments into oscillons or transients, the GWs frequencies can be reduced at the expense of the strength of the GW signal, still lying outside the scope of current GW detectors. 

As a fiducial model we considered the $\alpha$-attractor T-model for which the inflaton potential is symmetric and asymptotes to a constant value away from the central minimum. Our conclusions should apply to asymmetric potentials such as the ones in $\alpha$-attractor E-models \cite{Carrasco:2015pla,Carrasco:2015rva} as well as power-law Monodromy type potentials \cite{Silverstein:2008sg,McAllister:2014mpa}, since the dynamics of the oscillons and transients is determined mainly by the region near the central minimum.\footnote{Note that the actual embedding of the $\alpha$-attractor inflaton potentials in Supergravity suffers from instabilities in auxiliary fields for $M<\sqrt{2}\mpl$ \cite{Kallosh:2013yoa,Renaux-Petel:2015mga,Krajewski:2018moi,Turzynski:2018zup,Iarygina:2018kee}. We have ignored this theoretical constraint and considered the single-field models at the phenomenological level. However, to ensure the transition to a radiation dominated universe, i.e., the succesful completion of reheating, at some point one has to account for couplings of the inflaton oscillons to other fields. They eventually lead to the decay of the oscillons into relativistic matter \cite{Hertzberg:2010yz,Saffin:2016kof}, ending the matter-dominated state of expansion.}

In future work we plan to consider dynamical gravitational clustering, as well as self-gravitating individual objects.

\section{Acknowledgements}

We thank Eiichiro Komatsu and Philip Mocz for useful discussions. MA is supported by a DOE grant DE-SC0018216.


\bibliographystyle{apsrev}
\bibliography{bibGravityPertFromOscillonsAndTransientsAfterInflation}{}
\appendix
\section{Poisson Solver}
\label{sec:Poisson}

From our simulations we find the energy density field at each lattice cite, $\rho_{x,y,z}$, which acts as a source for the gravitational potential in the Poisson equation \eqref{eq:NonLinearPoisson}. To find $\Phi^{\rm nl}_{x,y,z}$, we then solve the Poisson equation with periodic boundary conditions. Since the Laplacian operator is linear, in (discrete) Fourier space eq. \eqref{eq:NonLinearPoisson} reduces to a simple algebraic form
\Beq
\tilde{\Phi}^{\rm nl}_{k_x,k_y,k_z}&=\frac{a^2\tilde{\rho}_{k_x,k_y,k_z}dx^2}{2\mpl^2\tilde{D}(k_x,k_y,k_z)}\,,\\
\tilde{D}(k_x,k_y,k_z)&=-6+2\!\!\sum\limits_{i=x,y,z}\!\!\cos\left(\frac{2\pi}{N} k_i\right)\,,
\Eeq
where each integer Fourier mode wavenumber can be $0\leq k_i<N$ (but $k_x^2+k_y^2+k_z^2\neq0$), with $N^3$ being the total number of lattice points in the real-space cubic lattice. Since we consider linear metric perturbations, we assume they have zero spatial mean, i.e., we put $\tilde{\Phi}^{\rm nl}_{k_x=0,k_y=0,k_z=0}=0$. The co-moving distance between neighboring lattice points, $dx$, is constant. Note also that as a co-moving (discrete) real-space Laplacian operator, we use the standard second-order, $\mathcal{O}(dx^2)$, accurate expression
\Beq
\nabla^2\Phi^{\rm nl}_{x,y,z}&=\frac{D[\Phi^{\rm nl}_{x,y,z}]}{dx^2}\,,\\
D[\Phi^{\rm nl}_{x,y,z}]&=\Phi^{\rm nl}_{x+1,y,z}+\Phi^{\rm nl}_{x-1,y,z}\\
&+\Phi^{\rm nl}_{x,y+1,z}+\Phi^{\rm nl}_{x,y-1,z}\\
&+\Phi^{\rm nl}_{x,y,z+1}+\Phi^{\rm nl}_{x,y,z-1}\\
&-6\Phi^{\rm nl}_{x,y,z}\,.
\Eeq
This is consistent with {\it LatticeEasy}, since it uses the same implementation to calculate the Laplacian of $\phi$.

We can also compute the magnitude of the gravitational acceleration, $g=|\boldsymbol{\nabla}\Phi^{\rm nl}|/a$. To calculate the co-moving gradient we again use a standard finite difference implementation
\Beq
\nabla_i\Phi^{\rm nl}_{x,y,z}&=\frac{G_i[\Phi^{\rm nl}_{x,y,z}]}{dx}\,,\\
G_x[\Phi^{\rm nl}_{x,y,z}]&=\Phi^{\rm nl}_{x,y,z}-\Phi^{\rm nl}_{x-1,y,z}\,,\\
G_y[\Phi^{\rm nl}_{x,y,z}]&=\Phi^{\rm nl}_{x,y,z}-\Phi^{\rm nl}_{x,y-1,z}\,,\\
G_z[\Phi^{\rm nl}_{x,y,z}]&=\Phi^{\rm nl}_{x,y,z}-\Phi^{\rm nl}_{x,y,z-1}\,.
\Eeq

\end{document}